\documentclass[aip,pop,reprint,letter,noshowpacs,floatfix]{revtex4-1}

\usepackage{graphicx}
\usepackage{color}

\newcommand{\dif}{\ensuremath{\mathrm{d}}}
\newcommand\etal{\emph{et al.}} 
\newcommand\xhat{\ensuremath{\mathbf{\hat{x}}}}
\newcommand\yhat{\ensuremath{\mathbf{\hat{y}}}}
\newcommand\zhat{\ensuremath{\mathbf{\hat{z}}}}
\newcommand\bhat{\ensuremath{\mathbf{\hat{b}}}}
\newcommand\xmax{\ensuremath{x_{\mathrm{max}}}}
\newcommand\zmax{\ensuremath{z_{\mathrm{max}}}}
\newcommand\eymax{\ensuremath{E_{y,\mathrm{max}}}}
\newcommand\xn{\ensuremath{\mathbf{x}_n}}

\begin{document}

\title{The plasmoid instability during asymmetric inflow magnetic
  reconnection}

\author{Nicholas A. Murphy} 

\email[Electronic mail: ]{namurphy@cfa.harvard.edu}

\affiliation{Harvard-Smithsonian Center for Astrophysics, Cambridge,
  Massachusetts 02138, USA}

\author{Aleida K. Young}

\affiliation{Harvard-Smithsonian Center for Astrophysics, 
  Cambridge, Massachusetts 02138, USA}

\affiliation{Department of Atmospheric, Oceanic and Space Sciences,
  University of Michigan, Ann Arbor, Michigan 48109, USA} 

\author{Chengcai Shen}

\affiliation{Harvard-Smithsonian Center for Astrophysics, Cambridge,
  Massachusetts 02138, USA}

\affiliation{Yunnan Astronomical Observatory, Chinese Academy of
  Sciences, Kunming, Yunnan 650011, China}

\affiliation{Graduate School of the Chinese Academy of Sciences,
  Beijing 100049, China} 

\author{Jun Lin}

\affiliation{Harvard-Smithsonian Center for Astrophysics, 
  Cambridge, Massachusetts 02138, USA}

\affiliation{Yunnan Astronomical Observatory, Chinese Academy of
  Sciences, Kunming, Yunnan 650011, China}

\author{Lei Ni}
\affiliation{Yunnan Astronomical Observatory, Chinese Academy of
  Sciences, Kunming, Yunnan 650011, China}

\begin{abstract}
  
  Theoretical studies of the plasmoid instability generally assume
  that the reconnecting magnetic fields are symmetric.  We relax this
  assumption by performing two-dimensional resistive
  magnetohydrodynamic simulations of the plasmoid instability during
  asymmetric inflow magnetic reconnection.  Magnetic asymmetry
  modifies the onset, scaling, and dynamics of this instability.
  Magnetic islands develop preferentially into the weak magnetic field
  upstream region.  Outflow jets from individual X-points impact
  plasmoids obliquely rather than directly as in the symmetric case.
  Consequently, deposition of momentum by the outflow jets into the
  plasmoids is less efficient, the plasmoids develop net vorticity,
  and shear flow slows down secondary merging between islands.
  Secondary merging events have asymmetry along both the inflow and
  outflow directions.  Downstream plasma is more turbulent in cases
  with magnetic asymmetry because islands are able to roll around each
  other after exiting the current sheet.  As in the symmetric case,
  plasmoid formation facilitates faster reconnection for at least
  small and moderate magnetic asymmetries.  However, when the upstream
  magnetic field strengths differ by a factor of four, the
  reconnection rate plateaus at a lower value than expected from
  scaling the symmetric results.  We perform a parameter study to
  investigate the onset of the plasmoid instability as a function of
  magnetic asymmetry and domain size.  There exist domain sizes for
  which symmetric simulations are stable but asymmetric simulations
  are unstable, suggesting that moderate magnetic asymmetry is
  somewhat destabilizing.  We discuss the implications for plasmoid
  and flux rope formation in solar eruptions, laboratory reconnection
  experiments, and space plasmas.  The differences between symmetric
  and asymmetric simulations provide some hints regarding the nature
  of the three-dimensional plasmoid instability.

\end{abstract}

\maketitle

\section{INTRODUCTION}

The Sweet-Parker model of magnetic reconnection\cite{sweet:1958,
  *parker:1957, *parker:1963} predicts the formation of very high
aspect ratio current sheets in solar and astrophysical plasmas.
However, these high Lundquist number current sheets are unstable to
the formation of plasmoids.\cite{loureiro:2007, baalrud:2011,
  baalrud:2012, ni:2010, bhattacharjee:2009, samtaney:2009,
  huang:2010:plasmoid, huang:2011, ni:2012A, ni:2012B, shepherd:2010,
  barta:2011A, shen:2011, mei:2012} This plasmoid instability leads to
significant departures from the classical view of laminar,
Sweet-Parker-like reconnection.

Like the tearing mode,\cite{furth:1963} the linear properties of the
plasmoid instability have been investigated analytically by performing
an asymptotic matching analysis over an appropriate choice of
equilibrium.\cite{loureiro:2007, baalrud:2011, baalrud:2012, ni:2010,
  bhattacharjee:2009} The linear growth rate scales as $S^{1/4}V_A/L$
while the number of plasmoids scales as $S^{3/8}$.  Here, $V_A$ is the
upstream Alfv\'en speed, $L$ is the half-length of the current sheet,
$\eta$ is the resistivity, and $S\equiv L V_A/\eta$ is the Lundquist
number.  Numerical tests have confirmed these growth rates and
determined the eigenmode structure.  \cite{samtaney:2009,
  huang:2010:plasmoid} That the growth rate scales as the Lundquist
number to a positive exponent is significant: Sweet-Parker-like
reconnection layers become more unstable with increasing Lundquist
number.  The positive exponent occurs in part because the thickness of
Sweet-Parker current sheets scales as $\delta \sim S^{-1/2}$.  In
contrast, the growth rate of the tearing instability scales as
$S^{-3/5}$ for the constant-$\psi$ regime and $S^{-1/3}$ for the
nonconstant-$\psi$ regime in slab geometry.\cite{coppi:1976} In Hall
MHD, the linear growth rate is enhanced when the ion inertial length
exceeds the resistive skin depth.\cite{baalrud:2011} In three
dimensions, oblique modes of the plasmoid instability may develop when
a guide field is present because the locations of rational surfaces
are not always the surface where the reconnecting component of the
magnetic field reverses.\cite{baalrud:2012}

Plasmoid formation has been shown to onset when the Lundquist number
of a current sheet exceeds a critical value, $S_c$.
\cite{biskamp:1986, bhattacharjee:2009, huang:2010:plasmoid, ni:2012A,
  shen:2011, baty:2012, loureiro:2012} The most commonly quoted value
for the critical Lundquist number is $S_c \sim 10^4$, but there is
considerable variation in the values found for $S_c$.  Bhattacharjee
\etal\cite{bhattacharjee:2009} and Huang
\etal\cite{huang:2010:plasmoid} find that $S_c$ is around $3\times
10^4$ or $4\times 10^4$, while Shen and coauthors\cite{shen:2011}
determine that $S_c\approx 900$ for a different configuration.  Ni
\etal\cite{ni:2012A} find that the onset criterion depends on the
upstream plasma $\beta$, denoted $\beta_0$.  They find that $S_c$ is
between $2000$ and $3000$ for $\beta_0=0.2$, but this increases to
between $8000$ and $1\times 10^4$ for $\beta_0=50$.  They use
isothermal initial conditions for most of their simulations, but find
that many of the differences are reduced when uniform density initial
conditions are used instead.  The range in values for $S_c$ indicates
that the onset of the plasmoid instability is not just a function of
the Lundquist number, but also depends on the configuration of the
problem and the basic plasma parameters.  The value (or range in
values) for $S_c$ is important because statistical models of plasmoids
\cite{uzdensky:2010, huang:2012:dist, loureiro:2012, fermo:2010} often
assume that individual plasmoids are separated by marginally stable
current sheets.

The nonlinear evolution of the plasmoid instability has been
investigated by several groups.\cite{bhattacharjee:2009,
samtaney:2009, huang:2010:plasmoid, huang:2011, ni:2012A, ni:2012B,
shepherd:2010, barta:2011A, shen:2011, mei:2012} Surprisingly,
two-dimensional, symmetric resistive MHD simulations have shown that
the dimensionless reconnection rate levels off at {$\sim$}$0.01$ for
$S\gtrsim S_c$. \cite{bhattacharjee:2009, huang:2010:plasmoid}
Reconnection is therefore fast (i.e., independent of the Lundquist
number).  Reconnection rates in solar flares typically range from
${\sim}${$0.001$} to {$\sim$}$0.1$,\cite{dere:1996, *jqiu:2002,
*isobe:2005, *jjing:2005, *Nagashima:2006, *Narukage:2006} and it
has been argued that the role of the plasmoid instability is to
trigger collisionless reconnection to allow reconnection rates closer
to $0.1$. \cite{shepherd:2010, huang:2011}

Most simulations of the plasmoid instability assume symmetric inflow.
This approach reduces computing time because high resolution is
required only near the symmetry axis and only half of the domain needs
to be evolved in time.  This simplifies the analysis because magnetic
and velocity nulls located along the symmetry axis are easy to find.
However, the reconnection process will in general have some asymmetry.
Asymmetric inflow reconnection occurs when the upstream magnetic field
strengths and/or densities differ.  \cite{labellehamer:1995,
  ugai:2000, cassak:asym, cassak:hall, cassak:dissipation,
  malakit:2010, borovsky:2007, birn:2008, *birn:2010,
  pritchett:2008:asym, pritchett:2009, tanaka:2008, tanaka:2010,
  aunai:2013, swisdak:2003} Such conditions occur at Earth's dayside
magnetopause \cite{phan:1996, *ku:1997, *paschmann:2013,
  hasegawa:2010} and magnetotail,\cite{oieroset:2004} during
turbulence,\cite{servidio:2009, *servidio:2010} in laboratory plasma
experiments,\cite{yamada:1997A, murphy:mrx, beidler:2011, rogers:1995}
and in the solar atmosphere.\cite{linton:2006A, nnakamura:2012,
  su:2013:prominence, murphy:double} Reconnection can also have
asymmetric outflow,\cite{murphy:retreat, murphy:asym, murphy:double,
  oka:2008, *reeves:2010A, *Galsgaard:2002} as well as
three-dimensional asymmetries. \cite{Katz:2010, *Egedal:2011,
  *AlHachami:2010, *Pontin:2011:review, *lukin:2011, *Dorfman:2013,
  *wyper:2013}

In this paper, we perform and analyze two-dimensional resistive MHD
simulations of the plasmoid instability during asymmetric inflow
magnetic reconnection.  In Section \ref{numerical}, we describe the
numerical method and problem setup.  In Section \ref{nonlinear}, we
present the simulation results and compare the nonlinear evolution of
this instability during symmetric and asymmetric reconnection.  In
Section \ref{onset}, we discuss how the onset criterion for this
instability varies with asymmetry.  Section \ref{consequences}
contains a discussion of the observational consequences of these
simulations for the solar atmosphere, laboratory plasmas, and the
magnetosphere.  Section \ref{discussion} contains our discussion and
conclusions.

\section{NUMERICAL METHOD AND PROBLEM SETUP\label{numerical}}

The NIMROD code\cite{sovinec:jcp, sovinec:2010} solves the equations
of extended MHD using a finite element representation for two
dimensions and a finite Fourier series expansion for the third
dimension.  In dimensionless form, the equations solved for the
two-dimensional simulations reported in this paper are
\begin{eqnarray}
  \frac{\partial \rho}{\partial t}
  + \nabla \cdot \left( \rho \mathbf{V} \right)
  = \nabla \cdot D \nabla \rho,  \label{continuity}
  \\
  \frac{\partial \mathbf{B}}{\partial t} 
  =
  - \nabla \times 
  \left(
    \eta \mathbf{J} - \mathbf{V}\times\mathbf{B}
  \right), \label{farohms}
  \\
  \mathbf{J} = \nabla \times \mathbf{B}, \label{ampere}
  \\
  \rho 
  \left(
    \frac{\partial \mathbf{V}}{\partial t}
    + \mathbf{V} \cdot \nabla \mathbf{V}
  \right)
  = 
  \mathbf{J}\times\mathbf{B}
  - \nabla{p}
  + \nabla \cdot \rho \nu \nabla \mathbf{V}, \label{momentum}
  \\
  \frac{\rho}{\gamma-1}
  \left(
    \frac{\partial T}{\partial t} + \mathbf{V} \cdot \nabla T
  \right)
  =
  - \frac{p}{2} \nabla \cdot \mathbf{V}
  - \nabla \cdot \mathbf{q}
  + Q, \label{temperature}
\end{eqnarray}
where the variables are given by: $\mathbf{B}$, magnetic field;
$\mathbf{V}$, bulk plasma velocity; $\mathbf{J}$, current density;
$\rho$, plasma density; $p$, plasma pressure; $T$, temperature; $\nu$,
kinematic viscosity; $D$, an artificial density diffusivity; and
$\gamma=5/3$, the ratio of specific heats.  The heat source term
includes Ohmic and viscous heating,
\begin{equation}
  Q = \eta J^2 + \nu \rho \nabla \mathbf{V}^T\!\!:\!\nabla\mathbf{V}.
\end{equation}
The heat flux vector is given by 
\begin{equation} 
  \mathbf{q}=-\rho \left[ \chi_\|\bhat\bhat + \chi_\perp\left(
    \mathbf{I} - \bhat\bhat \right) \right]\cdot\nabla T,
\end{equation}
where \bhat\ is a unit vector in the direction of the magnetic field.
The parallel and perpendicular thermal diffusivities are given by
$\chi_\|$ and $\chi_\perp$, respectively.  The diffusivities are
uniform and given by $\eta = 10^{-3}$, $\nu = 2\times 10^{-3}$, $D =
10^{-3}$, $\chi_\| = 10^{-2}$, and $\chi_\perp = 5\times 10^{-4}$.
The normalizations are identical to those presented in
Refs.\ \onlinecite{murphy:retreat, murphy:double}.  Divergence
cleaning is used to prevent the accumulation of divergence
error.\cite{sovinec:jcp}

The initial conditions consist of a perturbed asymmetric Harris sheet
with uniform density and no guide field.\cite{birn:2008, *birn:2010,
  murphy:double} We define \xhat\ as the outflow direction, \yhat\ as
the out-of-plane direction, and \zhat\ as the inflow direction.  The
initial equilibrium is given by
\begin{eqnarray}
  B_{x0}(z) & = &B_{R0}
  \left[
    \frac{\tanh\left(\frac{z}{\delta_0}-b\right)+b}{1+b}
  \right],
  \\
  p_(z) & = & \frac{1}{2} 
  \left(1-B_x^2\right) + \beta_{R0}\frac{B^2_{R0}}{2},
  \\
  \rho(z) & = & 1
\end{eqnarray}
Here, $b$ controls the asymmetry of the magnetic field, $\delta_0$ is
the initial thickness of the current sheet, and $\beta_{R0} \equiv
p_{R0}/\left(B^2_{R0}/2\right)$.  Throughout this paper, the
subscripts `L' and `R' correspond to the asymptotic initial amplitudes
of fields for $z<0$ and $z>0$, respectively, while the subscript `$0$'
corresponds to $t=0$.  The ratio of the asymptotic upstream magnetic
fields is given by
\begin{equation}
  R \equiv \frac{B_{L}}{B_{R}}, \label{R0}
\end{equation}
where we use the conventions that $B_L,B_R>0$ and $B_L\leq B_R$ such
that $0<R\leq 1$.  Because $R$ is a function of time, we compare
simulations using $R_0$ instead.  We take $B_{R0}=1$ and
$\beta_{R0}=1$ for all simulations.  Because $B_{R0}$ is kept constant
and $B_{L0}$ is decreased for asymmetric cases, it is important to
note that there is consequently less magnetic flux and energy
available during asymmetric simulations.  To maintain total pressure
balance, plasma pressure is higher on the weak magnetic field side
such that $p_{L0}>p_{R0}$ and $\beta_{L0}>\beta_{R0}$ when $R_0<1$.

Each simulation is seeded with initial magnetic perturbations of the
form
\begin{equation}
  \mathbf{B}_p(x,z) = \sum_{q=1}^{N_q} \nabla\times \left( A_q \yhat
  \right), \label{perturb_total}
\end{equation}
where
\begin{equation}
  A_q =
  -B_q\Delta\exp\left[- \left( \frac{x-x_q}{\Delta} \right)^2 - \left(
    \frac{z}{\Delta} \right)^2 \right] .
  \label{perturb_specific}
\end{equation}
Multiple perturbations are used so that the initial conditions are not
symmetric about $x=0$.  The simulations presented in Section
\ref{nonlinear} have stronger secondary initial perturbations than the
simulations in Section \ref{onset}.

The domain consists of $m_x$ by $m_z$ rectangular finite elements
along the outflow and inflow directions, respectively.  Sixth order
finite elements are used for all simulations.  The size of the
computational domain is given by $x \in [-\xmax,\xmax]$ and $z \in
[-\zmax,\zmax]$, where $\xmax$ varies between simulations and $\zmax =
16$ for all simulations.  We
model four different initial asymmetries: $R_0 \in \{0.125, 0.25, 0.5,
1.0 \}$ (with $R_0=0.125$ only considered in Section \ref{onset}).
For our largest simulations with $\xmax=150$, we use
$\left( m_x, m_z \right) = \left( 576, 32 \right)$ for $R_0=1$,
$\left( m_x, m_z \right) = \left( 336, 60 \right)$ for $R_0=0.5$, and
$\left( m_x, m_z \right) = \left( 306, 60 \right)$ for $R_0=0.25$.
We use no-slip, conducting wall boundaries for each upstream region.
The simulations are periodic along the outflow direction;
consequently, downstream pressure effects are likely to be more
important than in simulations with open outflow boundary conditions.

During simulations with $R_0=1$, significant mesh packing is required
over a relatively small portion of the computational domain.  During
simulations with $R_0\ne 1$, most X-points slowly drift into the
strong field upstream region.\cite{cassak:asym, murphy:double}
Consequently, high resolution along the inflow direction is required
over a larger portion of the domain.  Relatively high resolution is
required along the outflow direction for all plasmoid unstable
simulations so that secondary reconnection associated with island
merging can be sufficiently resolved.  Insufficient resolution often
yields spurious nulls as a result of numerical dispersion error (see
also Ref.\ \onlinecite{MWan:2013}).  The resolution requirements along
the outflow direction are most stringent for cases with $R_0=1$.
Higher resolution is required in the weak field upstream region than
in the strong field upstream region.  When initializing these
simulations, we chose to increase resolution in the current sheet
region to ensure convergence rather than increasing $\zmax$;
consequently, the late-time evolution of asymmetric simulations is
impacted somewhat by the conducting wall boundary on the weak magnetic
field side.

During the initial simulations performed for this study, we used
relatively large amplitude symmetric initial perturbations with a
characteristic length scale comparable to the size of the
computational domain.  Simulations starting from these initial
conditions showed plasma sloshing back and forth along the inflow
direction, suggesting that there was a large-scale pressure imbalance
in our initial conditions. As an analogy, consider two regions of
antiparallel magnetic field that are initially in total pressure
equilibrium but have different magnetic field amplitudes, $B_0$ and
$R_0B_0$ with $0<R_0<1$.  Then suppose that the magnetic field
strength in each region is decreased by $\delta B$.  The weak magnetic
field region will then have a total pressure that is
$\left(1-R_0\right)B_0\delta B$ greater than the strong magnetic field
side, resulting in a large-scale force imbalance.  We greatly reduced
the sloshing behavior by using localized, small amplitude initial
perturbations.

For symmetric simulations, it is straightforward to find magnetic and
velocity nulls located on the symmetry axis, $z=0$.  To determine the
null positions, we find neighboring grid points between which the
signs of $B_z$ or $V_x$ change and apply Brent's
method.\cite{brent:1973} Brent's method combines the robustness of
bisection and root bracketing for ill-behaved functions while using
inverse quadratic interpolation which converges quickly for
well-behaved functions.  The finite element basis functions are used
to interpolate between grid points.  Magnetic nulls are classified as
X-type if $\frac{\partial B_z}{\partial x} \frac{\partial
  B_x}{\partial z}>0$ and O-type if $\frac{\partial B_z}{\partial x}
\frac{\partial B_x}{\partial z}<0$.

For asymmetric simulations and to find nulls not located along the
symmetry axis during symmetric simulations, we combine several
techniques to find magnetic and velocity nulls.  First, we search for
grid cells with changes in sign for both components of the vector
field while excluding cells along the conducting wall
boundaries. Second, we use bilinear
interpolation\cite{haynes:2007:trilinear} to provide an initial
approximation for the null position.  We exclude cases where the
inferred location is outside the grid cell boundary.  Rarely, this
step finds multiple nulls within the same grid cell.  While bilinear
or trilinear interpolation is highly appropriate for simulations with
fields defined only at discrete locations, higher order accuracy can
be obtained for NIMROD simulations by interpolating the finite element
basis functions.  Our third step is to perform a few iterations of the
method of steepest descent on the magnitude of the vector field.  This
method is robust but converges slowly.  Fourth, we use Broyden's
method\cite{broyden:1965} to converge on the null position.  If
Broyden's method does not converge, we alternate between the third and
fourth steps until convergence is achieved.  To classify magnetic
nulls, we define a matrix $\mathbf{M}$ with elements $M_{ij} =
\partial B_i/\partial x_j$.  The null is X-type if
$\det{\mathbf{M}}<0$ and O-type if
$\det{\mathbf{M}}>0$.\cite{parnell:1996}

Finally, we define several quantities to facilitate our analysis and
comparisons between different simulations.  When investigating the
scaling of asymmetric inflow reconnection,
Refs.\ \onlinecite{cassak:asym, swisdak:2007} derived that the outflow
velocity scales as a hybrid Alfv\'en speed that is a function of the
magnetic field strengths and densities in both upstream regions,
\begin{equation}
  V_{Ah} \equiv \sqrt{
    \frac{B_LB_R\left(B_L+B_R\right)}{\rho_LB_R+\rho_RB_L}}. \label{Vah}
\end{equation}
This simplifies to $V_{Ah0}=\sqrt{R_0}$ when $B_{R0}=1$ and
$\rho_0=\rho_{L0}=\rho_{R0}=1$.  This corresponds to $V_{Ah0}=1$ for
$R_0=1$, $V_{Ah0} \approx 0.71$ for $R_0=0.5$, $V_{Ah0} = 0.5$ for
$R_0 = 0.25$, and $V_{Ah0} \approx 0.35$ for $R_0 = 0.125$.
Throughout this paper, we use the convention
\begin{equation}
  L \equiv \frac{2}{3}x_\mathrm{max}.
\end{equation}
to account for the current sheet not extending the entire distance
along the outflow direction.  Using these quantities, we define the
hybrid Lundquist number to be\cite{murphy:double}
\begin{equation}
  S_h \equiv \frac{L V_{Ah}}{\eta} . \label{Sh}
\end{equation}
The reconnection rate for collisional asymmetric reconnection without
plasmoids is predicted to be \cite{cassak:asym}
\begin{equation}
  E_\mathrm{predict} = \sqrt{\frac{\eta V_{Ah}}{L}B_LB_R}\label{epredict}.
\end{equation}
Using our conventions that $\rho_{L0}=\rho_{R0}=1$,
$R_0=B_{L0}/B_{R0}$, and $B_{R0}=1$, we find that
\begin{equation}
  E_{\mathrm{predict},0} = R_0^{3/4}     
  \eta^{1/2}L^{-1/2}
  .\label{epredict0}
\end{equation}
This expression is an exact result for our conventions and our use of
dimensionless parameters, but should not be considered a general
result.

To describe the dynamics of X-points and O-points, we define $\xn =
(x_n,z_n)$ as the position of a magnetic null.  The plasma flow
velocity at the null is $\mathbf{V}(\xn)$.  The time derivative of the
null's position is $\dif \xn / \dif t$.  In resistive MHD, differences
between $\mathbf{V}(\xn)$ and $\dif \xn / \dif t$ must be due to
resistive diffusion of the magnetic field.  In general, $\mathbf{V}(x)
\ne \dif \xn / \dif t$ when asymmetry is present and $\eta > 0$.  For
O-points, we have generally found that $\mathbf{V}(\xn) \approx \dif
\xn / \dif t$.  For X-points, the difference can be a non-negligible
fraction of the nearby Alfv\'en speed. \cite{murphy:retreat,
  murphy:double} During X-line retreat in resistive MHD,
Ref.\ \onlinecite{murphy:retreat} has shown that the physical
mechanism behind this difference is resistive diffusion of the inflow
component of the magnetic field along the inflow direction.

\section{SIMULATION RESULTS\label{nonlinear}}

In this section, we compare simulations of the plasmoid instability
during symmetric inflow reconnection and asymmetric inflow
reconnection.  

\subsection{Nonlinear Dynamics}

\begin{figure*}
  \begin{center}
    \vspace{-0.4cm}
    \hspace{-0.9cm}
   \includegraphics[scale=1.10]{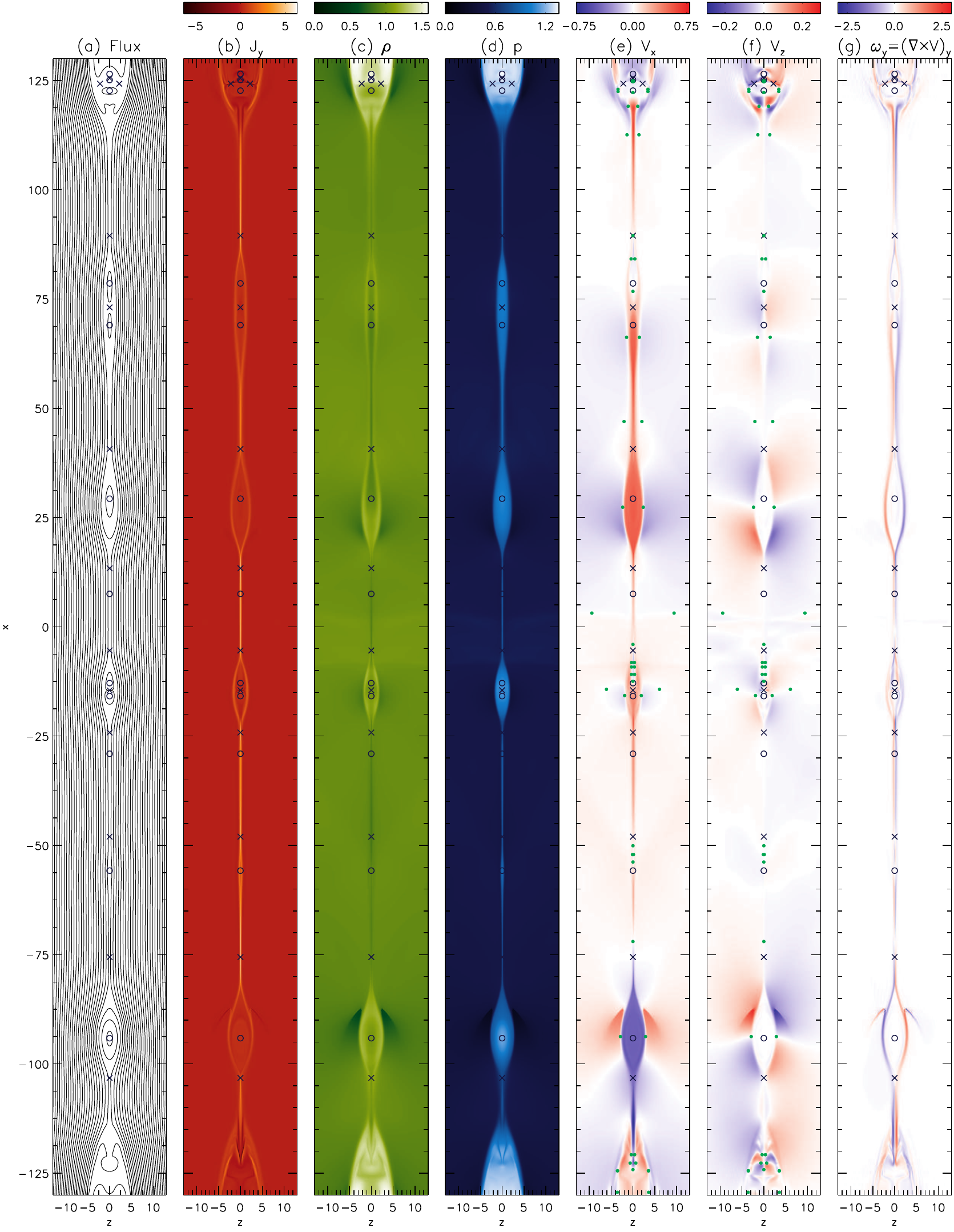}
  \end{center}
  \vspace{-0.75cm}
  \caption{ Simulation of the plasmoid instability with symmetric
    inflow ($R_0=1$) at $t=324$. Shown are contours of the (a)
    magnetic flux; (b) out-of-plane current density, $J_y$; (c) plasma
    density, $\rho$; (d) plasma pressure, $p$; (e) the outflow
    component of velocity, $V_x$; (f) the inflow component of
    velocity, $V_z$; and (g) the vorticity, $\omega_y$.  X-points are
    denoted by `$\times$' and O-points are denoted by `$\circ$'.  The
    green dots in panels (e) and (f) represent flow stagnation points.
    \label{paperplot_sym}}
\end{figure*}

\begin{figure*}
  \begin{center}
    \vspace{-0.4cm}
    \hspace{-0.9cm}
    \includegraphics[scale=1.10]{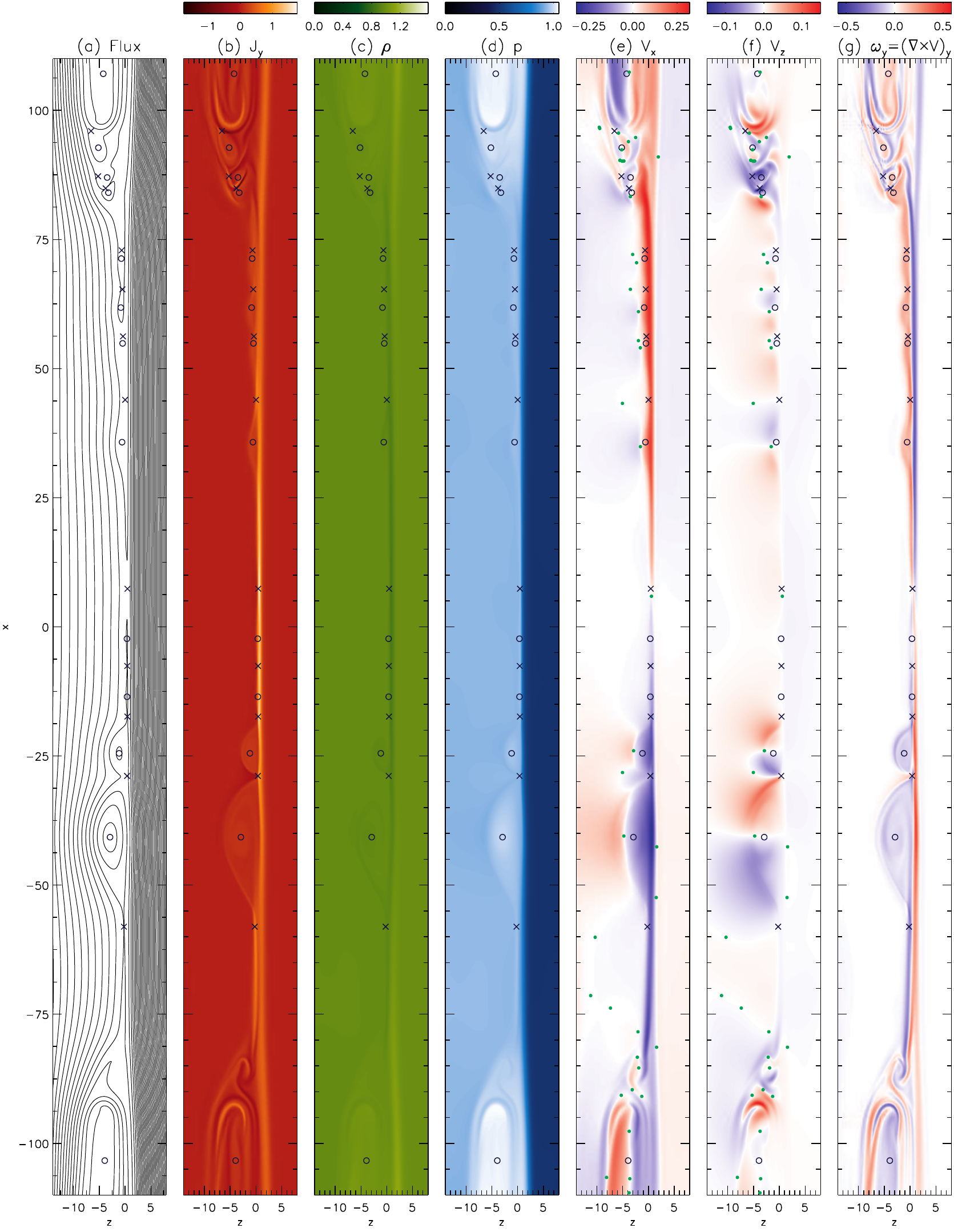}
  \end{center}
  \vspace{-0.75cm}
  \caption{ Simulation of the plasmoid instability with asymmetric
    inflow ($R_0=0.25$) at $t=731$. Shown are contours of the (a)
    magnetic flux; (b) out-of-plane current density, $J_y$; (c) plasma
    density, $\rho$; (d) plasma pressure, $p$; (e) the outflow
    component of velocity, $V_x$; (f) the inflow component of
    velocity, $V_z$; and (g) the vorticity, $\omega_y$.  X-points are
    denoted by `$\times$' and O-points are denoted by `$\circ$'.  The
    green dots in panels (e) and (f) represent flow stagnation points.
    \label{paperplot_asym}}
\end{figure*}

During the plasmoid instability with symmetric inflow
(Fig.\ \ref{paperplot_sym}), the X-points and O-points within the
current sheet region are located along $z=0$ because of symmetry.
Some X-points and O-points develop away from $z=0$ in the downstream
region away from the current sheet.  Within the current sheet, the
outflow jets impact each island directly so that momentum from the
jets is transported efficiently into the islands.  The islands do not
develop net vorticity [Fig.\ \ref{paperplot_sym}(g)].  The quadrupole
structure in $V_z$ around large islands corresponds to upstream plasma
being temporarily pushed out of the way as the island moves along with
the reconnection outflow; this also corresponds to reverse flow in
$V_x$ in the upstream region as plasma flows back to fill the area
behind the moving island.  The flow patterns in the outflow region
show some degree of order because of the assumption of symmetry, and
thus are not best described as being turbulent
[Figs.\ \ref{paperplot_sym}(e) and \ref{paperplot_sym}(f);
  cf.\ Ref.\ \onlinecite{mei:2012}].  There exist several flow
stagnation points (in the simulation reference frame) along $z=0$ as
well as symmetrically in both upstream regions.  Often, flow
stagnation points along the symmetry axis are located between an
X-point and a more central maximum in plasma pressure.  This occurs
because flow stagnation points preferentially occur where pressure
gradient and tension forces cancel.\cite{murphy:asym, murphy:retreat}
Individual X-points are often located in proximity to a neighboring
magnetic island so that the small-scale reconnection has asymmetric
outflow.  The plasma flow velocity at each individual X-point differs
somewhat from the velocity of each X-point [$\mathbf{V}(\xn) \ne \dif
  \xn / \dif t$] such that there is net plasma flow across each
X-point.\cite{murphy:retreat}

There exist many qualitative differences between the symmetric and
asymmetric plasmoid instability (Fig.\ \ref{paperplot_asym}).  In
contrast to the symmetric case, the positions along the inflow
direction of the X-points and O-points in the current sheet are not
constrained to $z=0$.  The X-points generally drift slowly into the
strong field upstream region. \cite{cassak:asym, murphy:double} In
general, the X-points closer to $x=0$ are displaced more into the
strong field upstream region than X-points in the periphery.  When
islands merge with each other or the large island downstream of the
reconnection region, X-points sometimes end up being displaced into
the weak field upstream region.  Because of field line rigidity, very
little happens in the strong field upstream region.

The most apparent feature of the plasmoid instability during
asymmetric inflow reconnection is that the plasmoids develop
preferentially into the weak field upstream region
[Fig.\ \ref{paperplot_asym}(a)].  The effect has been noted in prior
simulations of asymmetric inflow reconnection without secondary island
formation.\cite{ugai:2000, cassak:asym, pritchett:2008:asym,
  murphy:double} Reconnection outflow jets therefore impact islands
obliquely rather than directly.  This has two main consequences for
the structure and dynamics of individual plasmoids.  First, the
islands develop net vorticity as seen in
Fig.\ \ref{paperplot_asym}(g).  This effect has been observed in prior
simulations of line-tied asymmetric reconnection.\cite{murphy:double}
Second, momentum transport from the reconnection outflow jets into the
islands is less efficient.  The reconnection outflow is able to
partially bypass the islands.  Consequently, the islands that form
during asymmetric inflow reconnection have slower velocities relative
to $V_{Ah0}$.  For the largest cases with $\xmax = 150$, islands
propagate at velocities of $\lesssim 0.5 V_{A0}$ for $R_0=1$ and at
velocities of $\lesssim 0.3 V_{Ah0}$ for $R_0=0.25$.  Islands that
develop in asymmetric simulations therefore remain in the current
sheet for significantly longer.

Secondary merging events occur during both symmetric and asymmetric
simulations when two islands reconnect with each other to form a
single island.  The flux contained in the resulting island equals the
greater of the two initial fluxes.\cite{fermo:2010} An example of
secondary merging during a symmetric simulation is the X-point at
$x=-14.5$ between two neighboring magnetic islands in Figure
\ref{paperplot_sym} (see also the X-point at $x=73.1$ for an earlier
stage of secondary merging).  During symmetric simulations, the
outflow from secondary reconnection events is symmetric, but the
inflow is generally asymmetric because merging islands typically have
different sizes.  The secondary outflow jets are impeded symmetrically
by the upstream magnetic field.  Prior scaling studies show that
symmetric obstructions reduce the reconnection rate much more
effectively than when only one outflow jet is
obstructed. \cite{murphy:asym} However, because there is no freedom
for plasmoids to roll around each other, secondary reconnection is
able to be more effective.

\begin{figure}
  \begin{center}
    \includegraphics[scale=1.03]{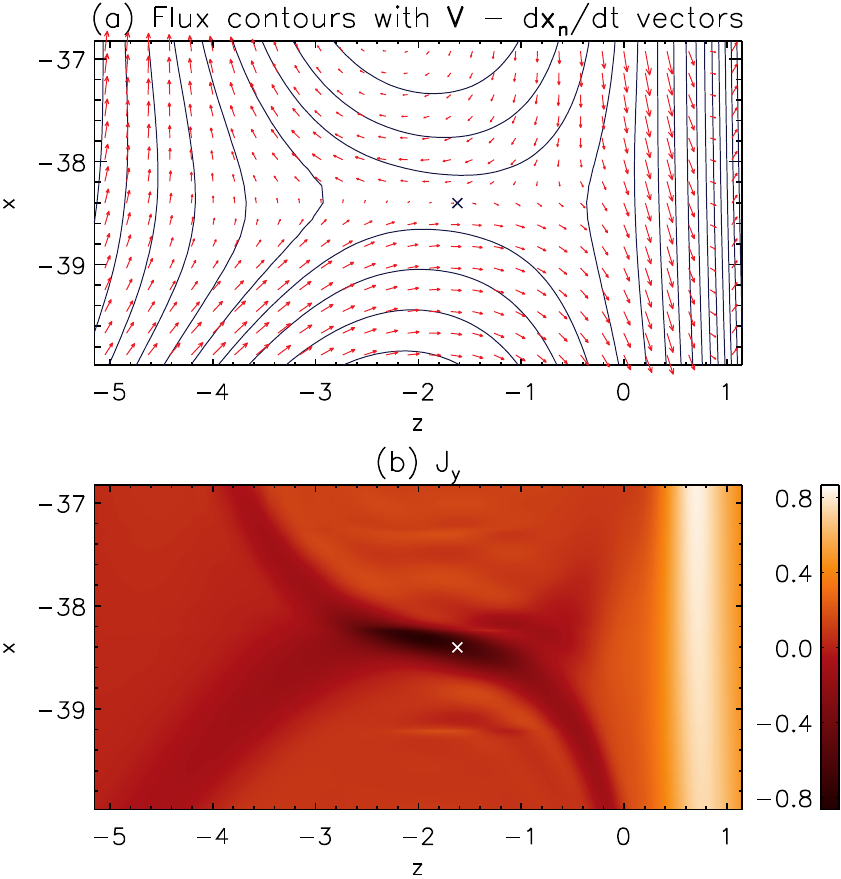}
  \end{center}
  \caption{
    \label{paperplot_secondary}
    Secondary merging between two plasmoids at $t=818$ for $R_0=0.25$
    and $\xmax=150$.  Shown are (a) magnetic flux contours and
    velocity vectors in the frame of the moving X-point, $\mathbf{V} -
    \dif \mathbf{x}_n/\dif t$, and (b) out-of-plane current density,
    $J_y$.  The local flow pattern is dominated by shear flow
    associated with island vorticity.  The longest vector corresponds
    to a velocity of $0.17$ while the characteristic Alfv\'en speed in
    the islands is {$\sim$}$0.1$\@.  The X-point is denoted by
    `$\times$'.    }
\end{figure}

During simulations with $R_0 \ne 1$, the reconnection process has
asymmetry along both the inflow and outflow directions.  An instance
of secondary merging late in the simulation with $\xmax=150$ and
$R_0=0.25$ is shown in Fig.\ \ref{paperplot_secondary}.  The
characteristic magnetic field strength is $0.1$ in the island, which
corresponds to a local upstream Alfv\'en speed of $0.1$ given that
$\rho\approx 1$.  At this time, $\mathbf{x}_n = \left( -38.40 , -1.62
\right)$, $\mathbf{V}(\mathbf{x}_n) = \left(-0.100 , -0.018\right)$,
and $\dif \mathbf{x}_n/\dif t=\left(-0.098,-0.040\right)$; this
indicates that the net plasma flow across the X-point along the $z$
direction is {$\sim$}$0.22$ of the local upstream Alfv\'en speed.  The
tilting of the current sheet occurs because the island below is larger
than the island above.

The most important feature of this secondary reconnection is the flow
pattern shown in Fig.\ \ref{paperplot_secondary}(a) in the reference
frame of the moving X-point.  The characteristic inflow/outflow
pattern typically associated with two-dimensional resistive
reconnection is absent.  In contrast, the flow pattern in the current
sheet region is dominated by shear flow associated with the vorticity
in the islands.  The magnitude of the shear flow is comparable to the
local upstream Alfv\'en speed (rather than the hybrid Alfv\'en speed),
which has been shown to suppress and slow down the reconnection
process. \cite{XLChen:1990, *QChen:1997, *LaHaye:2010, *JHLi:2010,
  *cassak:2011A, *cassak:2011B} The merging process began around
$t\approx 765$ [when the X-point between these two islands first had
$J_y(\xn)<0$] and was not quite complete by the end of the simulation
at $t=1021$.  This velocity shear does not occur during secondary
merging in symmetric simulations because those islands lack net
vorticity.

\subsection{Reconnection Rate}

\begin{figure}
  \includegraphics[scale=1.04]{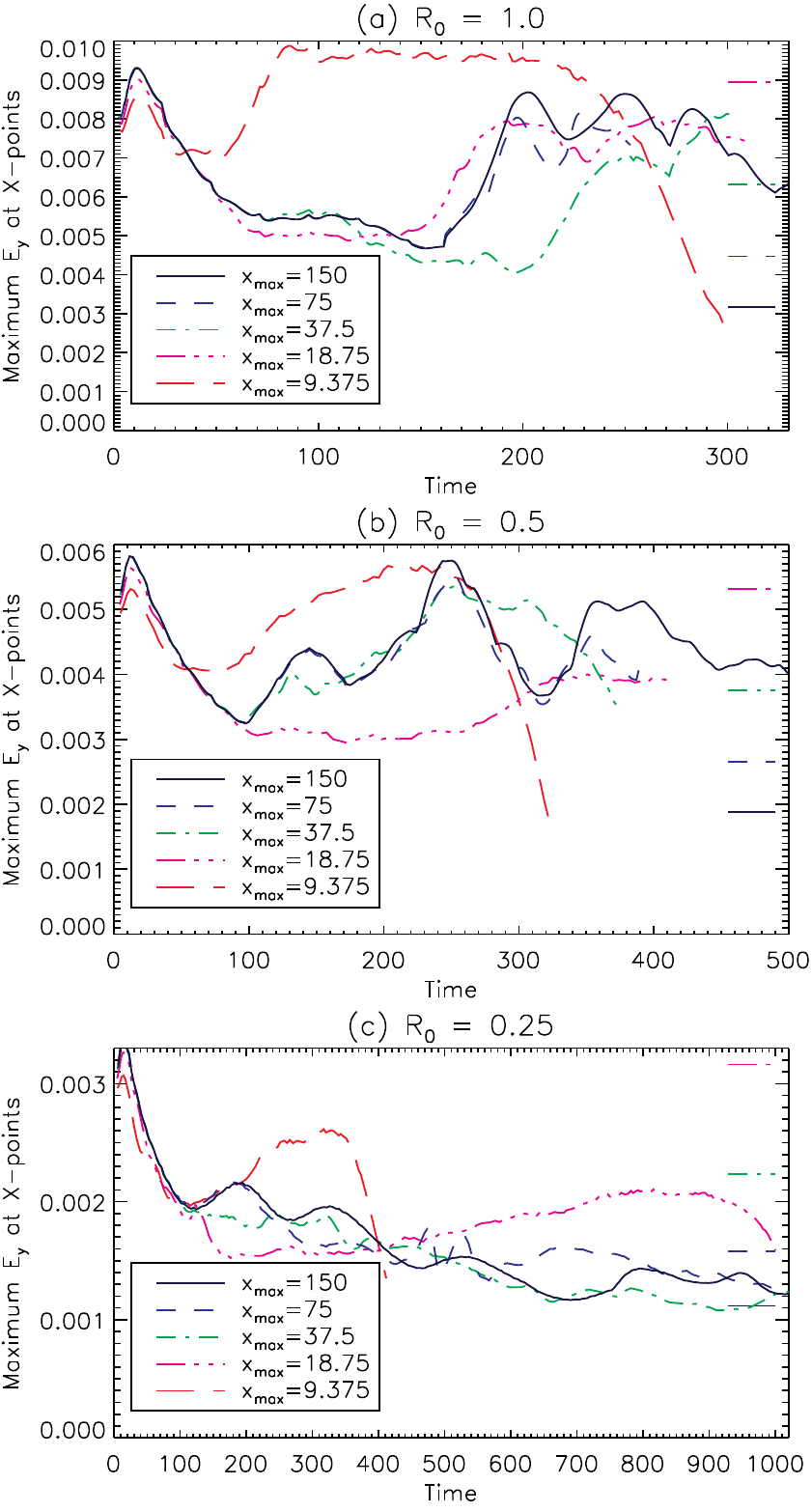}
  \caption{ The reconnection rate as a function of time for
    simulations with different asymmetries and domain sizes.  The
    reconnection rate is given by the maximum out-of-plane electric
    field among all of the X-points in a simulation. The horizontal
    line segments indicate the reconnection rate predicted by
    Eq.\ \ref{epredict0}.  The predictions for $\xmax = 9.375$ are not
    shown, but are factors of $\sqrt{2}$ greater than the predictions
    for $\xmax=18.75$.}
    \label{paperplot_ey}
\end{figure}

A key result from two-dimensional resistive MHD simulations of the
symmetric plasmoid instability is that the reconnection rate becomes
insensitive to the Lundquist number for $S>S_c$.  We now investigate
how the reconnection rate is impacted by magnetic asymmetry during the
plasmoid instability.

We define the reconnection rate, $\eymax$, to be the maximum
out-of-plane electric field among all X-points within the current
sheet, $-L \leq x \leq L$.  This directly represents the amount of
flux being reconnected at the most active X-point.  An alternative
method used for symmetric simulations is to compare the inflow
velocity to the outflow velocity; however, this results in some
ambiguity for asymmetric inflow simulations because X-points drift at
different velocities along the inflow direction, and the inflow
velocities differ in each upstream region.

Fig.\ \ref{paperplot_ey} shows the reconnection rate over time for
simulations with different magnetic asymmetries, $R_0 \in \{ 0.25,
0.5, 1.0 \}$, and different domain sizes, $\xmax \in \{ 9.375, 18.75,
37.5, 75, 150 \}$.  Plasmoids develop for all simulations with $\xmax
\ge 37.5$, but do not develop or are not important for simulations
with $\xmax \le 18.75$.  At $t=0$, $\eymax$ is large because the
initial current sheet is thin, so we concentrate our analysis on $t
\gtrsim 75/V_{Ah0}$ after reconnection develops.

We use two strategies to compare the reconnection rates between
different simulations.  The first strategy is to compare small and
large simulations directly, without considering predicted values for
the reconnection rate.  Nominally, the effects not included in
theoretical models such as viscosity, thermal conduction, and
downstream pressure will have similar consequences in each simulation.
The second strategy is to compare $\eymax$ with the predicted value
given by Eq.\ \ref{epredict0}.  This method allows us to determine the
extent to which additional factors are responsible for slowing down
the reconnection rate.

In symmetric simulations [Fig.\ \ref{paperplot_ey}(a)], the peak
reconnection rate is only weakly sensitive to changes in the domain
size (and therefore Lundquist number).  The peak reconnection rate for
$\xmax = 9.375$ is comparable to the peak reconnection rate for
$\xmax=150$.  For cases without plasmoids, Eq.\ \ref{epredict0}
overpredicts the reconnection rate.  This indicates that effects not
included in the derivation of this equation slow down reconnection.
Overall, this result corroborates the reconnection rate enhancement
due to the plasmoid instability found in previous
works. \cite{bhattacharjee:2009, huang:2010:plasmoid, shen:2011,
  loureiro:2012} The reconnection rate levels off at $\eymax \sim
0.008$.

For $R_0=0.5$ [Fig.\ \ref{paperplot_ey}(b)], the results are
qualitatively similar to $R_0=1$.  The peak reconnection rates are
comparable among most simulations.  The exception is $\xmax=18.75$
which has a slower reconnection rate than the rest of the simulations.
The plasmoid instability continues to show enhancement of the
reconnection rate for this asymmetry.  The reconnection rate levels
off between about $0.004$ and $0.005$, which is consistent with what
one would expect from the symmetric results and the scaling for our
simulation setup that $\eymax \propto R_0^{3/4}$.

There are some qualitative differences for $R_0=0.25$
[Fig.\ \ref{paperplot_ey}(c)].  The peak reconnection rates for the
cases without plasmoids are higher than the cases with plasmoids.
However, the reconnection rates for the three cases with plasmoids are
comparable, indicating that the reconnection rate is leveling off as
the domain size is increased. Therefore, we conclude that there
remains enhancement of the reconnection rate due to the plasmoid
instability.  However, when we compare to the predicted values from
Eq.\ \ref{epredict0}, the reconnection rates from the simulation are
generally much lower.  For the largest two domain sizes, the
prediction becomes comparable to the simulation results.  The
reconnection rate levels off at $\eymax$ between about $0.0012$ and
$0.0015$, compared to the value of $0.0028$ expected from scaling the
symmetric case.

Overall, we conclude that the reconnection rate is still enhanced due
to the plasmoid instability at least for asymmetries of $0.25 \lesssim
R_0 \leq 1$.  However, for $R_0 = 0.25$, the reconnection rate levels
off at a lower value than would be inferred from the symmetric
results.  For low to moderate asymmetries, the enhancement in the
reconnection rate is comparable to what would be expected from scaling
the symmetric case.

\section{ONSET OF INSTABILITY\label{onset}}

\begin{figure}
  \includegraphics[scale=1.04]{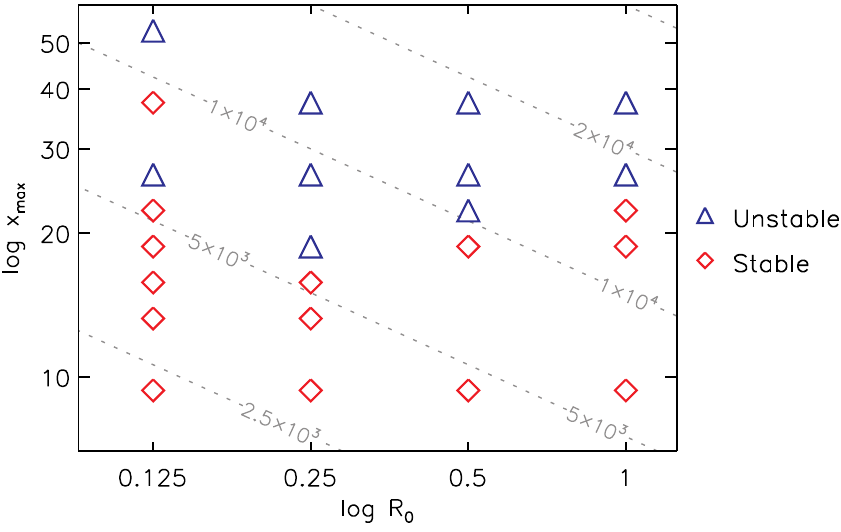}
  \caption{ Results from a parameter study to test the dependence of
    the initial magnetic asymmetry ratio, $R_0 \in \{0.125, 0.25,
    0.50, 1.00\}$, and the half-size of the computational domain along
    the outflow direction, \xmax, on the onset of the plasmoid
    instability.  Simulations are classified as unstable if new
    X-points form in the current sheet before $t = 5\xmax/V_{Ah0}$.
    Blue triangles indicate instability, while red diamonds indicate
    stability.  The gray dotted lines are contours of constant initial
    hybrid Lundquist number, $S_{h0}$.
  \label{paperplot_parameterspace}}
\end{figure}

In this section, we perform a grid of nonlinear simulations with
different domain sizes and magnetic asymmetries to investigate the
onset criterion of the plasmoid instability during symmetric and
asymmetric inflow magnetic reconnection.  We vary the initial magnetic
asymmetry, $R_0$, and the size of the computational domain along the
outflow direction, \xmax\@.  We place a single perturbation at the
origin, and a much smaller perturbation off to one side to allow
outflow asymmetry.  Simulations are classified as unstable if new
X-points appear in the current sheet region before $t=5\xmax/V_{Ah0}$.
If no new X-points form, then the simulations are classified as
stable.  For a given asymmetry, the smallest unstable simulation
typically yields only short-lived islands that exit the current sheet
and merge with the main outflow island soon after formation.  The
results of this parameter study are shown in
Fig.\ \ref{paperplot_parameterspace}.

For symmetric cases, we find the critical Lundquist number to be $S_c
\approx 1.6 \times 10^4$.  This is somewhat smaller than the Huang
\etal \cite{huang:2010:plasmoid} value of $S_c \sim 4\times 10^4$,
comparable to Loureiro \etal\cite{loureiro:2012} value of $S_c\sim 1.2
\times 10^4$, and larger than the values reported by Ni
\etal\cite{ni:2012A} and Shen \etal\cite{shen:2011} The differences
between these values occur in part because the onset criterion depends
on the configuration and plasma parameters as well as the Lundquist
number.

For asymmetric cases, we initially hypothesized that plasmoids form
when the hybrid Lundquist number exceeds a critical value, $S_{ch}$,
which is a constant and not a function of asymmetry.  This hypothesis
predicts system sizes for which the symmetric case is unstable and
asymmetric cases are stable, and that the demarcation between stable
and unstable configurations are aligned with a contour of constant
$S_{h0}$.  However, Fig.\ \ref{paperplot_parameterspace} indicates
that this hypothesis is incorrect.

Figure \ref{paperplot_parameterspace} shows that plasmoids form for
smaller domain sizes when there is a moderate magnetic asymmetry, $R_0
\in \{0.25,0.5\}$, than when there is symmetry, $R_0=1$.  For
$R_0=1.0$, $0.5$, $0.25$, and $0.125$, the smallest unstable domain
sizes are $\xmax = 26.52$, $22.30$, $18.75$, and $26.52$,
respectively.  In contrast to our hypothesis, moderate magnetic
asymmetry is somewhat destabilizing for the plasmoid instability.

Anomalously, the simulation with $R_0=0.125$ and $\xmax=37.5$ does not
show plasmoid formation out to $t=7\xmax/V_{Ah0}$ despite new X-lines
forming in a smaller simulation with $\xmax = 26.52$.  New X-lines do
appear for $\xmax = 53.03$.  When the simulations with $R_0=0.125$
were repeated with different initial perturbations, none of them
showed evidence for plasmoid formation for $\xmax \leq 37.5$.  These
results suggest that strong magnetic asymmetry ($R_0 \lesssim 0.125$)
is stabilizing.

We now consider possibilities for why moderate magnetic asymmetry
cases show plasmoid formation for shorter domain sizes than symmetric
cases.  Our first hypothesis was that there was simply inadequate time
for plasmoids in the symmetric case to form.  To test this, we ran the
largest stable symmetric simulation (with $R_0=1$ and $\xmax=22.30$)
twice as long. No new X-points developed in the extended run of this
simulation, so this hypothesis is not supported.  Our second
hypothesis was that the difference in the flow properties led to the
difference in the onset domain size.  This could be caused by either
the outflow being slower such that there is more time for
instabilities to develop, or that there will be less shear flow
stabilization for cases with slower outflow.  We test this hypothesis
by repeating the largest stable and smallest unstable simulations for
different asymmetries with $\mathrm{Pm}=0.5$ and $\mathrm{Pm}=8$
instead of $\mathrm{Pm}=2$ (where $\mathrm{Pm} = \nu/\eta$ is the
magnetic Prandtl number).  In contrast to the prediction of this
hypothesis, higher viscosity leads to larger domain sizes becoming
stable.  This hypothesis is not supported; however, an important
caveat is that changing the viscosity affects the structure of the
diffusion region as well as slowing down the outflow.  The results do
not rule out an alternative hypothesis that the development of the
instability is correlated with intermittency in the flow.

The results from this parameter study do not explain why simulations
with moderate asymmetry are unstable for smaller domain sizes than
symmetric cases, or why strongly asymmetric simulations are more
stable.  Some insight may be gained by performing a linear stability
analysis for the plasmoid instability during asymmetric inflow
reconnection, including an asymptotic matching analysis and linear
simulations,\cite{dubois:1979, *einaudi:1984, *ding:1992, furth:1963,
  loureiro:2007, ni:2010} to see how the growth rate and eigenmode
structure are modified by magnetic asymmetry.

\section{OBSERVATIONAL CONSEQUENCES\label{consequences}}

In this section, we make observational predictions for the behavior of
plasmoids in solar, laboratory, and space plasmas during asymmetric
inflow reconnection.  Differences between simulation and observation
will provide key insights into the roles of important effects not
included in the simulations, such as collisionless effects and
three-dimensional effects.

\subsection{Solar Atmosphere}

The standard model of solar eruptions predicts the formation of an
elongated current sheet in the wake behind the rising flux rope.
\cite{jlin:2000} While unambiguous identification of these structures
is difficult without magnetic field information, several features
classified as current sheets have been observed in the solar corona.
\cite{ko:2003,*takasao:2012, ciaravella:2002, *webb:2003, *jlin:2005,
  *ciaravella:2008, *schettino:2010, *savage:2010, *reeves:2011,
  *landi:2012, *savage:2012B, *ciaravella:2013} These current sheets
necessarily have asymmetric outflow and may have asymmetric inflow.
\cite{murphy:double, su:2013:prominence} Large blobs are frequently
observed during these events \cite{ko:2003,*takasao:2012} which have
been interpreted to be the result of a tearing or plasmoid
instability.\cite{Riley:2007, *jLin:2007} `Monster plasmoids' such as
those predicted by Ref.\ \onlinecite{uzdensky:2010} are the most
likely features to be observed.

In principle, an offset of blobs toward the weak field upstream region
could be seen with instruments such as the Atmospheric Imaging
Assembly (AIA) on the Solar Dynamics Observatory, the X-Ray Telescope
on Hinode, and the Large Angle and Spectrometric Coronagraph (LASCO)
on the Solar and Heliospheric Observatory.  However, if the current
sheet is observed at a small angle with respect to the line of sight,
then blobs that are physically offset might not show an apparent
offset due to projection effects.  Constraints on the overall geometry
are therefore vital.  High-resolution, high-cadence observations by
AIA have been used to investigate vorticity in current sheet features
at low heights, \cite{mckenzie:2013} and it may be possible to find
evidence for vorticity in current sheet blobs.  However, the time
cadence of LASCO observations is too slow to observe vorticity on the
scales of individual blobs.

Asymmetric reconnection also occurs during solar jets,
\cite{nnakamura:2012} which happen when newly emerged flux reconnects
with pre-existing, overlying flux. \cite{yokoyama:1996, *shibata:2007}
The pre-existing flux is usually weaker, so our simulations predict
plasmoid development preferentially into that region as well as the
associated vortical motions.  However, the spatial and temporal
resolution requirements to observe small-scale structure and vortical
motions during jet reconnection are likely beyond the capabilities of
current instrumentation.

\subsection{Laboratory Experiments}

The current generation of dedicated laboratory experiments on magnetic
reconnection are stable or marginally stable against the formation of
multiple plasmoids or flux ropes. \cite{ji:2011} There is hope that
the next generation of reconnection experiments will have $S>S_c$ to
allow the plasmoid instability to be investigated in the laboratory.
If the geometry is toroidal or cylindrical, then the reconnection
process will necessarily have asymmetric inflow unless the radius is
very large.  In the Magnetic Reconnection Experiment (MRX),
reconnection is driven by ramping up or down currents carried within
toroidal flux cores.  During the `pull' mode of operation (see Fig.\ 5
of Ref.\ \onlinecite{yamada:1997A}), the low radius side of the
current sheet has stronger magnetic field than the high radius
upstream region, and the current sheet gradually drifts toward lower
radii.\cite{yamada:2011B, lukin:2003, murphy:mrx}

It is therefore likely that the effects discussed in this paper will
play some role in the dynamics of the plasmoid instability in future
experiments.  The asymmetry is not expected to be large, so the
effects are not likely to be as pronounced as in the simulations.  For
future experiments with configurations similar to MRX but with
$S>S_c$, our simulations predict the development of flux ropes
preferentially into the high radius upstream region and and the
development of net vorticity.  The current sheet is expected to drift
toward lower radii where the magnetic field is stronger.  Secondary
reconnection is expected to be less efficient than predicted by
two-dimensional, symmetric simulations because of the freedom of flux
ropes to roll around each other, especially in the downstream region.
However, these predictions are likely to be modified due to
cylindrical geometry effects not included in our paper.  Simulations
of the plasmoid instability in an experimental geometry will allow
direct comparisons to be made (see also Refs.\ \onlinecite{murphy:mrx,
  lukin:2003, dorfman:2008, lukin:2001}).

\subsection{Space Plasmas}

Plasmoids and flux ropes are frequently observed \emph{in situ} during
reconnection events in planetary magnetospheres and the solar wind.
Reconnection events at the dayside magnetopause and its flanks have
asymmetric inflow. \cite{phan:1996, *ku:1997, *paschmann:2013,
  hasegawa:2012} In contrast to our two-dimensional resistive MHD
simulations with an antiparallel magnetic field configuration,
reconnection in near-Earth space plasmas is in the collisionless
regime. \cite{oieroset:2001, oieroset:2004, mozer:2008} There is often
a guide field, and shear flow effects are often important for dayside
reconnection. \cite{labellehamer:1995} There is also density asymmetry
along with magnetic asymmetry.

For these reasons, we anticipate that there will be significant
qualitative differences between our simulations and \emph{in situ}
observations of asymmetric reconnection at the dayside magnetopause.
Nevertheless, some of the macroscopic features of our simulations may
remain applicable during collisionless asymmetric inflow reconnection.
Island development into the weak field upstream region is in principal
observable; pragmatically, this will be difficult to diagnose because
of the limited number of spacecraft taking measurements.  More
promising are measurements of vorticity in secondary flux ropes formed
by reconnection (see Ref.\ \onlinecite{hasegawa:2012} and references
therein), but care will be needed to distinguish these features from
Kelvin-Helmholtz vortices.  To allow more direct comparisons to
\emph{in situ} observations, it will be important to extend recently
performed two-fluid, \cite{cassak:hall, cassak:dissipation} hybrid,
\cite{aunai:2013} and fully kinetic\cite{swisdak:2003,
  pritchett:2008:asym, pritchett:2009, malakit:2010, tanaka:2010,
  aunai:2013} simulations of asymmetric inflow reconnection to
investigate the dynamics of secondary islands and flux ropes.

\section{DISCUSSION\label{discussion}}

In this paper, we perform resistive MHD simulations of the plasmoid
instability during magnetic reconnection with asymmetric upstream
magnetic fields.  This is in contrast to most studies of the plasmoid
instability which assume that the reconnection process is symmetric.
Relaxing the assumption of symmetric inflow leads to qualitatively
different results.

During symmetric simulations, the X-points and O-points within the
current sheet region are located along the symmetry axis so that
momentum from the outflow jets is efficiently transported into the
islands.  Secondary islands within the current sheet do not develop
net vorticity and are advected quickly out of the current sheet.
Secondary reconnection events associated with island merging have
symmetric outflow but asymmetric inflow.  In the downstream region,
some X-points and O-points are located away from the symmetry axis.
Because of the assumption of symmetry, the flow pattern in the
downstream region is structured and therefore not best described as
being turbulent.

During asymmetric simulations, the locations of the X-points and
O-points are offset from each other along the inflow direction.  There
is generally slow drifting of the current sheet into the strong field
upstream region.  X-points near the center of the sheet are displaced
further into the strong field upstream region than X-points near the
current sheet exits.  Not much happens in the strong field upstream
region because of field line rigidity.  Islands develop preferentially
into the weak field upstream region.  Consequently, outflow jets
impact the islands obliquely rather than directly so that net
vorticity develops and momentum transport into the islands is less
efficient.  During secondary reconnection, shear flow associated with
vorticity in the merging islands slows down reconnection.  The
downstream regions develop a complicated flow structure indicative of
turbulence, with islands able to roll around each other as well as
merge.

We compare the reconnection rate enhancement between symmetric and
asymmetric cases.  During high Lundquist number symmetric simulations,
the reconnection rate is enhanced above the value predicted from the
Sweet-Parker model.  This is consistent with previous
work.\cite{bhattacharjee:2009, huang:2010:plasmoid, shen:2011,
  loureiro:2012} During simulations with a magnetic asymmetry of
$R_0=0.5$, the reconnection rate enhancement is comparable to what is
expected from scaling the symmetric simulations.  For stronger
asymmetry ($R_0=0.25$), the reconnection rate is still enhanced by the
presence of plasmoids in the current sheet, but to less of a degree
than for $R_0 \in \{ 0.5, 1.0 \}$.  Scaling from the symmetric
simulations to $R_0=0.25$ cases overpredicts the reconnection rate.

We perform an onset study of the plasmoid instability for both
symmetric and asymmetric inflow reconnection.  For the symmetric case,
we find that the critical Lundquist number for the onset of the
instability is $S_c \approx 1.6 \times 10^4$.  This is consistent with
the canonical result that onset occurs when the Lundquist number is of
order $10^4$\@.  Interestingly, we find that there exist domain sizes
for which the symmetric case is stable but moderate asymmetry cases
(e.g., $R_0 \in \{0.25,0.5 \}$) are unstable.  This suggests that
moderate asymmetry has a destabilizing influence on the formation of
plasmoids.  However, our results indicate that strong asymmetry (e.g.,
$R_0 \lesssim 0.125$) has a stabilizing influence.

We predict observational consequences for the asymmetric plasmoid
instability in the solar atmosphere, laboratory experiments, and space
plasmas.  The most likely characteristics to be observed are the
development of islands preferentially into the weak field upstream
region and vorticity in the plasmoids.  In the solar atmosphere, high
spatial and temporal resolution will be required and projection
effects will make interpretation difficult.  Future laboratory
experiments may naturally have asymmetric inflow due to cylindrical
geometry effects, although the asymmetry may be relatively modest.
\emph{In situ} observations of flux ropes formed during asymmetric
inflow reconnection at Earth's dayside magnetopause provide an
opportunity to test predictions based on vorticity of secondary
islands; however, collisionless effects not included in our
simulations are expected to be important.  For all of these
situations, discrepancies between simulations and observations will
provide insight into the roles of three-dimensional and collisionless
effects.

Our simulations of the plasmoid instability during asymmetric inflow
reconnection hint at the nature of this instability in three
dimensions.  In particular, outflow jets from individual, small-scale
reconnection sites are more likely to impact flux ropes obliquely
rather than directly.  This may cause momentum transport from the jets
into the flux ropes to be less efficient and vorticity in the flux
ropes to develop.  Vorticity in flux ropes may lead to shear flow
stabilization when flux ropes merge.  Consequently, secondary merging
of flux ropes formed by the plasmoid instability should not
necessarily be assumed to be efficient (see also
Ref.\ \onlinecite{intrator:2009, *sun:2010}).  As in the downstream
region of our asymmetric simulations, flux ropes will likely have
additional freedom to roll around each other while merging or instead
of merging.  However, further study of the three-dimensional plasmoid
instability will be required before solid conclusions can be drawn.

There remain several open questions and promising areas of future work
regarding the plasmoid instability during both symmetric and
asymmetric inflow magnetic reconnection.  Thus far, no analytical
theory exists that describes the linear properties of the plasmoid
instability during asymmetric inflow reconnection; however, the
effects of asymmetry on the tearing mode have been
investigated.\cite{dubois:1979, *einaudi:1984, *ding:1992} The role of
collisionless effects and the transition to collisionless reconnection
have been investigated for the symmetric plasmoid instability
\cite{shepherd:2010, huang:2011} but not for the asymmetric case.  The
role of three-dimensional effects during the plasmoid instability
requires further study.  Additional investigation of the role of a
guide field will provide more insight into real situations, especially
for three-dimensional and fully kinetic simulations.  Statistical
models of magnetic islands in current sheets \cite{uzdensky:2010,
  fermo:2010, loureiro:2012, huang:2012:dist} will need to be extended
in order to incorporate asymmetry and three-dimensional effects.

\begin{acknowledgments}
  The authors thank A.\ Bhattacharjee, P.~A.\ Cassak, T.~G.\ Forbes,
  L.\ Guo, Y.-M.\ Huang, K.~E.\ Korreck, N.~F.\ Loureiro, M.\ Oka,
  J.~C.\ Raymond, K.~K.\ Reeves, S.~L.\ Savage, L.~S.\ Shepherd,
  C.~R.\ Sovinec, and H.~D.\ Winter for useful discussions.  The
  authors thank members of the NIMROD team for ongoing code
  development that helped make this work possible.  In particular, we
  thank J.~King for suggesting memory management strategies that
  allowed larger simulations to be performed.  Resources supporting
  this work were provided by the NASA High-End Computing (HEC) Program
  through the NASA Advanced Supercomputing (NAS) Division at Ames
  Research Center.  This research has benefited greatly from the use
  of NASA's Astrophysics Data Service.

  This research was supported by NASA grants NNX09AB17G, NNX11AB61G,
  and NNX12AB25G; NASA contract NNM07AB07C; and NSF SHINE grant
  AGS-1156076 to the Smithsonian Astrophysical Observatory.
  A.K.Y.\ acknowledges support from the NSF-REU solar physics program
  at the Center for Astrophysics, grant number ATM-0851866\@.

\end{acknowledgments}

\end{document}